\documentclass[aps,prl,reprint,superscriptaddress]{revtex4-2}

\usepackage{graphicx}
\usepackage{hyperref}

\begin{document}

\title{Cold Beam Optical Clock with Multifrequency Spectroscopy}
\author{William G. Tobias}
\email{william.g.tobias.ctr@us.navy.mil}
\affiliation{Computational Physics, Inc., 8001 Braddock Road, Suite 210, Springfield, VA 22151}
\affiliation{Precise Time Department, United States Naval Observatory, 3450 Massachusetts Avenue NW, Washington, DC 20392}
\author{Bryan Hemingway}
\author{Steven Peil}
\email{steven.e.peil.civ@us.navy.mil}
\affiliation{Precise Time Department, United States Naval Observatory, 3450 Massachusetts Avenue NW, Washington, DC 20392}
\date{\today}
\begin{abstract}
 We demonstrate an optical clock based on Ramsey-Bordé interferometry in a laser-cooled ${}^{40}$Ca beam. The mean velocity is reduced by an order of magnitude relative to a thermal beam and the transverse temperature approaches the Doppler limit, enabling the measurement of sub-kHz linewidth fringes in a compact interferometer. Using tailored phase and intensity modulation of the spectroscopy laser to add uniform frequency sidebands, we interrogate atoms throughout the transverse velocity distribution, increasing the Ramsey-Bordé fringe amplitude by a factor of 14 and improving the Allan deviation to $3.4\times 10^{-15}$ at one second averaging time. 
\end{abstract}
\maketitle

Ensembles of microwave atomic clocks form the backbone of worldwide timing infrastructure, achieving high stability over decades of continuous operation \cite{Wynands2005,Peil2017}. The development of optical clocks, which have demonstrated stabilities and uncertainties far better than their microwave counterparts, will stimulate future advances in timekeeping \cite{Ludlow2015}.

Optical clocks come in many forms, each representing trade-offs among stability, complexity, and availability. The highest-stability clocks are based on trapped ions or ultracold atoms in optical lattices \cite{Brewer2019, Aeppli2024}, and require multiple laser wavelengths, dead time for sample preparation, and frequent user intervention. This reduces the maximum practical clock availability \cite{Lodewyck2016, Baynham2018, Kobayashi2020}. A continuous output can be maintained using microwave or optical flywheel oscillators but this necessarily degrades the stability~\cite{Grebing2016,Hachisu2018,Yao2019,Milner2019,Formichella2024}. Relatively few co-trapped atoms can be interrogated simultaneously since interaction-induced frequency shifts increase with density; the resulting quantum projection noise (QPN) sets a classical limit on trapped-atom clock performance.

Clocks based on thermal atomic or molecular vapors \cite{Martin2018,Roslund2024,Ahern2024} or atomic beams \cite{Kersten1999,McFerran2010,Shang2017, Olson2019} enable continuous spectroscopy of orders-of-magnitude larger atomic ensembles, with correspondingly decreased QPN. Relative to trapped-atom clocks, however, the instability is increased by broad optical transition linewidths, frequency drift, and high atomic velocities, which necessitate stringent control over laser alignment and source temperature to mitigate Doppler shifts \cite{Hemingway2020, Strathearn2023, Barger1981}. Frequency shifts can be reduced using apertures or counterpropagating spectroscopy lasers, methods which sample only a small part of the thermal velocity distribution.

Building on the recent demonstration of a thermal beam optical clock with instability below $10^{-15}$ \cite{Olson2019}, Ref.~\cite{Shang2022} proposed engineering a multifrequency laser spectrum to match the distribution of Doppler-shifted clock transitions and efficiently interrogate atoms at all velocities. This method is a coherent analogue of ``white-light" slowing \cite{Hoffnagle1988, Zhu1991}, where a broadband laser interacts resonantly with all atoms in a thermal beam, and is related to multiphoton direct frequency comb spectroscopy \cite{Kielpinski2006,Jayich2016}, in that many excitation pathways contribute constructively in probing a narrow transition.

Using multifrequency Ramsey-Bordé interferometry in a laser-cooled ${}^{40}$Ca beam, we demonstrate an optical clock that achieves short-term stability competitive with trapped-atom clocks in a low-complexity architecture specialized to continuous, long-term operation. Laser cooling narrows the clock transition linewidth and reduces the uncertainty associated with Doppler shifts while requiring no additional laser wavelengths relative to previous experiments \cite{McFerran2010,Olson2019,Hemingway2020}. Multifrequency spectroscopy enables parallel addressing and readout of an order-of-magnitude higher atom number, improving clock performance and illustrating a general-purpose technique for efficient narrow-line spectroscopy of atomic sources with broad velocity distributions.

\begin{figure}
    \centering
    \includegraphics[scale=1]{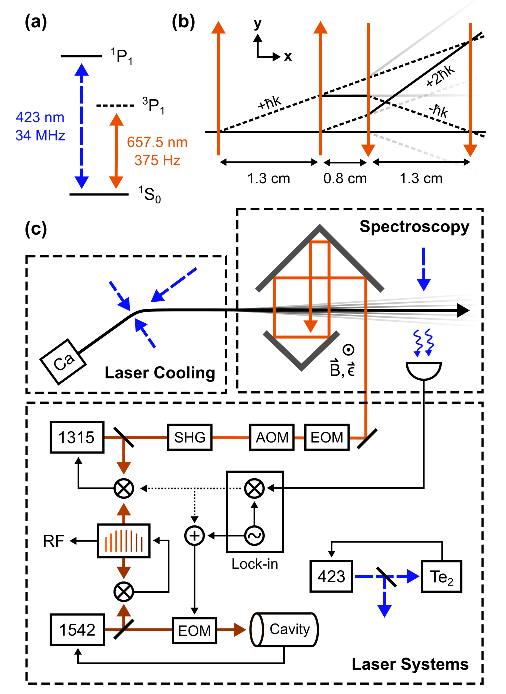}
    \caption{(a) ${}^{40}$Ca energy level diagram. (b) Ramsey-Bordé interferometry. Two pairs of counterpropagating lasers split the atomic wavefunction into ground state (solid line) and excited state (dashed line) components, producing two closed trajectories. Select lines are labeled with the total momentum transferred to each atom. (c) Experimental setup. Thermal atoms are emitted from an oven, laser-cooled, and deflected into the interferometer. The spectroscopy laser (1315~nm, doubled to 657.5~nm) is pre-stabilized to a frequency comb locked to a cavity-stabilized laser (1542~nm), with the atomic signal measured by lock-in detection and fed back to the laser frequency. The cooling and probe laser (423~nm) is stabilized to a Te${}_2$ vapor cell. SHG: second-harmonic generation; AOM: acousto-optic modulator; EOM: electro-optic modulator.}
    \label{fig:fig1}
\end{figure}

Clock spectroscopy in ${}^{40}$Ca is performed on the ${{}^1S_0-{}^3P_1}$ intercombination transition (Fig.~\ref{fig:fig1}(a)), the absolute frequency and systematic shifts of which have been characterized in prior research on optical frequency standards \cite{Morinaga1989,Oates2000,Degenhardt2005,Wilpers2006}.
Relative to other alkaline earth elements, the natural linewidth of 375 Hz is particularly well-suited to spectroscopy in an atomic beam: it is both broad enough to require only modest laser power and stabilization and narrow enough to enable improved stability relative to existing microwave references. An additional laser on the ${}^1S_0-{}^1P_1$ cycling transition is used for laser cooling and for probing the atomic state distribution following the interferometer \cite{McFerran2010,Hemingway2020}.

Ramsey-Bordé interferometry is used to generate narrow clock transition linewidths in the atomic beam \cite{Borde1984}. In this method, atoms experience four subsequent phase-coherent $\pi/2$ pulses from two pairs of counterpropagating laser beams (Fig.~\ref{fig:fig1}(b)). The atomic wavefunction splits with the absorption and emission of photons. Interference fringes associated with two recoil components---closed space-time trajectories with fixed total momentum exchanged between each atom and the optical fields---are produced at the output of the interferometer. The fringe period, and consequently the effective linewidth, depends on the total dark time elapsed in the first and last free-evolution regions. No relative phase is accumulated in the central region, where a single energy and momentum state connects to each recoil component at the output. The time dependence of the pulse intensity derives from the laser geometry and the velocity of the atoms; this differs from trapped-atom interferometers, where time-varying laser pulses are applied to stationary atoms \cite{Oates2000,Wilpers2002,Degenhardt2005,Wilpers2006}.

Figure \ref{fig:fig1}(c) shows the atomic and laser systems comprising the optical clock. Thermal ${}^{40}$Ca is produced in the ${}^1S_0$ ground state in an oven at 540~${}^\circ$C and is collimated using mechanical apertures. The velocity components parallel and perpendicular to the atomic beam are addressed using Zeeman slowing and two-dimensional magneto-optical trapping, detuned from the ${}^1S_0-{}^1P_1$ resonance at 423~nm by $-1020$~MHz and $-40$~MHz, respectively. The final cooling stage is oriented at an angle relative to the thermal beam, resulting in a slowed atom flux of $5(4)\times10^{10}$~s${}^{-1}$ being deflected towards the interferometer \cite{Witte1992}. While the average first-order Doppler shift is eliminated by the use of counterpropagating laser beams, the distribution of non-zero transverse velocities (perpendicular to the atomic beam) reduces the fraction of atoms resonant with the spectroscopy laser.

The interferometer is formed by two corner cube retroreflectors, oriented so that an input laser beam intersects the atomic beam at four locations separated by 1.3, 0.8, and 1.3~cm. To isolate the $m_J=0$ sublevel in the excited state, a small quantization magnetic field $|\vec B|=1.5$~G is applied perpendicular to the atomic and laser beams and parallel to the laser polarization $\vec \epsilon$. Following the interferometer, the atomic state is probed using lock-in detection with a time constant of 10 ms.

The spectroscopy wavelength $\lambda=657.5$~nm is produced by a frequency-doubled 1315 nm external cavity diode laser (ECDL), pre-stabilized to an optical cavity by the frequency chain illustrated in Fig.~\ref{fig:fig1}(c). A 1542~nm fiber laser is phase-modulated with an electro-optic modulator (EOM) to produce frequency sidebands with a controllable detuning from the carrier, one of which is locked by Pound-Drever-Hall spectroscopy to a high-finesse optical cavity. The carrier-sideband offset is additionally modulated for lock-in detection. A frequency comb is phase-locked to the 1542~nm laser carrier and the 1315~nm ECDL is phase-locked to the comb and then subsequently frequency doubled using second harmonic generation (SHG), transferring the short-term cavity stability to the spectroscopy laser. Frequency corrections derived from lock-in demodulation of the atomic spectroscopy signal are fed back to either the offset frequency between the 1542~nm laser and cavity or between the 1315~nm laser and comb (dotted lines, Fig.~\ref{fig:fig1}(c)). To produce the cooling and probe light, a 423~nm ECDL is stabilized by modulation transfer spectroscopy to a Te$_2$ vapor cell \cite{Taylor2018}. Additional lasers are phase-locked to this reference laser and frequency-shifted using acousto-optic modulators (AOMs) to generate the necessary detunings.

\begin{figure}
    \centering
    \includegraphics[scale=0.9]{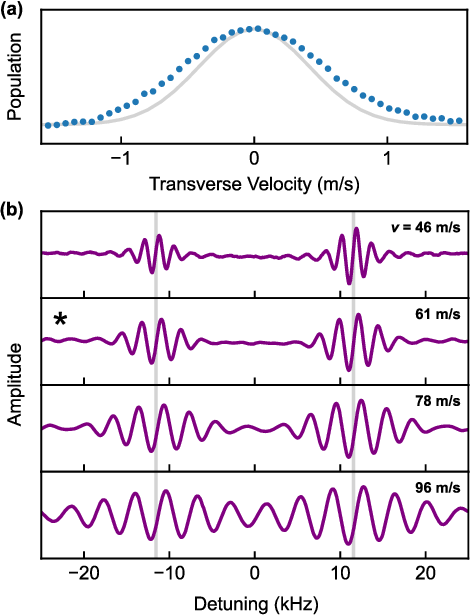}
    \caption{Characterizing the laser-cooled beam. (a) The measured one-dimensional transverse velocity distribution (blue points) corresponds to a temperature of $1.37$~mK, approaching the Doppler limit (gray line). (b) Ramsey-Bordé fringes measured over a range of longitudinal velocities $v$, scaled to a constant amplitude. Both recoil components are resolved (centered at the gray lines). The starred panel represents the experimental conditions for Figs. \ref{fig:fig3} and \ref{fig:fig4}.}
    \label{fig:fig2}
\end{figure}

To measure the transverse velocity distribution, we block the spectroscopy laser after a single pass through the atomic beam and measure the excited state population as a function of laser detuning (Fig.~\ref{fig:fig2}(a)). Laser frequency is converted to transverse velocity by the factor $k/2\pi=1/\lambda=1.52$~MHz/(m/s). Since there are no sub-Doppler cooling mechanisms present, the Doppler limit of $\hbar \gamma/2k_B = 0.81$~mK sets the minimum temperature achievable with laser cooling, where $\hbar$ is the reduced Planck constant, $\gamma=2\pi\times 34$~MHz is the natural linewidth, and $k_B$ is the Boltzmann constant. The measured temperature is $1.37$~mK.

The average longitudinal velocity $v$ (parallel to the atomic beam) is related to the Ramsey-Bordé fringe linewidth $\delta\nu$ as $\delta\nu=1/2t=v/4d$, where $t$ is the total dark time and $d=1.3$~cm is the distance between atom-laser interactions. We measure fringes at four Zeeman slower detunings equally spaced between $-980$ and $-1100$~MHz, and observe $\delta\nu=880\text{--}1840$~Hz, corresponding to $v=46\text{--}96$~m/s (Fig.~\ref{fig:fig2}(b)). Both recoil components are resolved, separated by twice the photon recoil frequency of $\hbar k^2/2m=2\pi\times11.56$~kHz, where $m$ is the atomic mass. The amplitude of the lower-frequency component, associated with the upper branch in Fig.~\ref{fig:fig1}(b), is reduced by spontaneous decay of the excited state in the central region of the interferometer, most prominently in the slowest atomic beam.

The optimal parameters for clock spectroscopy are determined by competition between the atomic quality factor and decoherence due to excited state decay, which both increase with the dark time $t$ \cite{Santarelli1999,Barry2020}. Additionally, the atomic flux is correlated with $t$ since varying the Zeeman slower detuning changes both the atomic velocity and slowed atom population. We find that the performance is optimized at $\delta\nu \approx 1.2$~kHz ($v=61$~m/s; starred panel, Fig.~\ref{fig:fig2}(b)).

In contrast to clocks where motion is nearly eliminated by trapping and cooling, Doppler shifts from transverse motion in atomic beams can substantially exceed the transit-time-limited spectral width of the spectroscopy pulses. This problem is exacerbated in a slowed beam because the spectral width is proportional to longitudinal velocity. The transverse velocity distribution has a frequency full-width-at-half-maximum of 1.96 MHz and the spectral width is less than 30~kHz, so only a small fraction of atoms are resonant with the spectroscopy laser. Reducing the laser beam waist increases the spectral width but is ultimately limited by wavefront inhomogeneity and practical difficulties in aligning diverging beams \cite{Strathearn2023, Olson2019}.

To address atoms at all transverse velocities, we modify the laser spectrum to add uniform frequency sidebands centered around the clock transition \cite{Shang2022}. The sideband detunings relative to the carrier are $p \nu_m$ for positive and negative integers $p$, where $\nu_m$ is the modulation frequency and the sideband amplitudes are set by the laser modulation parameters. In the first two atom-laser interactions, an atom with transverse velocity $v_T$ interacts with the sideband satisfying $kv_T\approx 2\pi\times p \nu_m$, where the sideband detuning compensates the Doppler shift. In the second two interactions, where the laser propagation direction is reversed relative to $v_T$, the same atom interacts with the opposite sideband ($p\rightarrow -p$). When $\nu_m$ is large relative to the spectral width of the spectroscopy pulses, each transverse velocity class interacts resonantly with at most a single pair of sidebands and produces fringes that add constructively. Generating sidebands with the same power as the carrier ensures uniform laser pulse area for all transverse velocities.

\begin{figure*}
    \centering
    \includegraphics[scale=0.9]{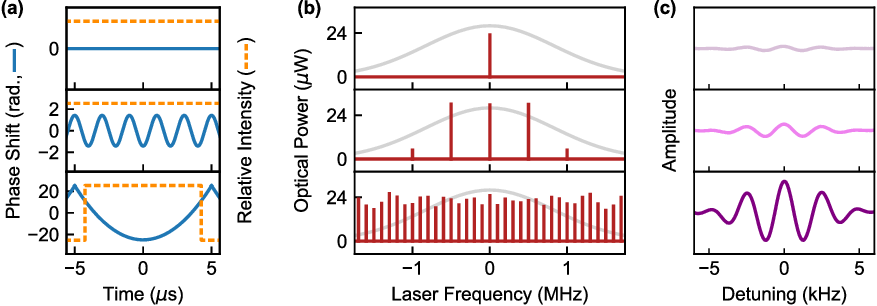}
    \caption{Single frequency, triple frequency, and multifrequency spectroscopy (SF, TF, MF; top, middle, and bottom rows). (a) Tailored phase (solid, blue) and intensity (dashed, orange) modulation waveforms. For TF spectroscopy, the phase is modulated sinusoidally. For MF spectroscopy, the phase is modulated quadratically and the intensity is chopped. (b) Optical spectra measured by self-heterodyne detection. Modulation produces approximately equal-power frequency components that span the distribution of transverse-velocity Doppler shifts (gray curve). (c) Measured Ramsey-Bordé fringes.}
    \label{fig:fig3}
\end{figure*}

We demonstrate two methods for modulating the laser spectrum using an EOM to control the laser phase and an AOM to control the laser intensity (Fig.~\ref{fig:fig3}; see also Fig.~\ref{fig:fig1}(c)).
Modulating only the phase sinusoidally with a peak-to-peak depth of $0.92\pi$~radians produces three equal-power lines separated by $\nu_m$. To generate many orders, we use tailored modulation of both phase and intensity, following previous work on electro-optic frequency comb generation \cite{Torres-Company2008,Wu2010}. Imprinting quadratic phase shifts on the pulses in a periodic train produces a flat, periodic frequency spectrum by time-to-frequency conversion, analogous to the position-to-momentum conversion produced by a lens \cite{Kolner1989,Azana2004}.
We modulate the phase quadratically with a depth of $15.9\pi$~radians and chop the intensity in-phase with full contrast and an $85\%$ duty cycle.  This produces a broad spectrum of laser lines spanning the transverse velocity distribution with optical powers varying by less than $\pm 15\%$.

Spectroscopy performance is determined by the overlap between the modulated laser spectrum and the distribution of transverse-velocity Doppler shifts as well as the relative magnitude of $\nu_m$ and the spectral width of the laser pulses. At small $\nu_m$, the density of laser lines is high but neighboring lines can interact with the same atoms, leading to crosstalk and reduced atomic coherence \cite{Shang2022}. We optimize $\nu_m$, the modulation depth, and the intensity chopping duty cycle by varying each parameter and measuring the fringe amplitudes. The modulation waveforms (a), measured laser spectra (b), and resultant fringes (c) are shown in Fig.~\ref{fig:fig3} for each modulation condition: no modulation (``single frequency", or SF), phase-only modulation at $\nu_m=500$~kHz (``triple frequency", TF), and simultaneous phase and intensity modulation at $\nu_m=100$~kHz (``multifrequency", MF).  TF and MF spectroscopy increase the fringe amplitudes by factors of 2.9 and 14.0, respectively. Though TF spectroscopy produces a smaller enhancement factor, it could be implemented in many laser systems with no added optical components using existing frequency modulation pathways.

For normal clock operation, the spectroscopy laser detuning is maintained at a zero crossing of the demodulated Ramsey-Bordé fringe and frequency corrections are added to the offset frequency between the 1542~nm laser and cavity (Fig.~\ref{fig:fig1}(c)), with the result that the spectroscopy laser and frequency comb track the atomic transition. The fractional frequency stability of the comb radiofrequency (RF) output at 10 MHz can then be compared to stable microwave sources. At short averaging times $\tau$, however, the available microwave references and measurement systems are less stable than the optical clock. To accurately measure the stability, the feedback path is modified and frequency corrections are added instead to the offset frequency between the 1315~nm laser and comb. In this geometry, the cavity and comb are independent of the spectroscopy laser at times sufficiently longer than the time constant of feedback from the atomic spectroscopy, and the offset frequency can be measured to determine the optical frequency difference between the optical clock and cavity. We estimate the noise floor of measurements at optical frequencies to be less than $2.3\times 10^{-15}$ at $1\le\tau\le10$~s based on comparisons of the optical cavity to additional cavities and microwave references.

\begin{figure}
    \centering
    \includegraphics[scale=0.9]{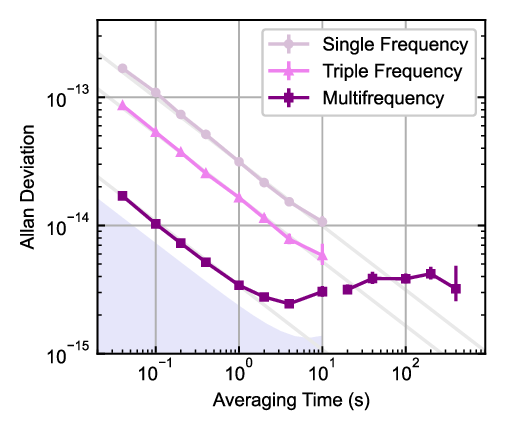}
    \caption{Optical clock performance. The clock is compared to an optical cavity at averaging times $\tau\le10$~s and to a hydrogen maser and CSO at $\tau>10$~s. Solid lines illustrate performance at short averaging times for the three methods of laser frequency modulation and the shaded region represents the estimated noise floor for $\tau\le10$~s. Error bars are 95\% confidence intervals.}
    \label{fig:fig4}
\end{figure}

We measure the Allan deviation of the ${}^{40}$Ca optical clock (Fig.~\ref{fig:fig4}).
For $\tau\le10$~s, the spectroscopy laser is compared to the optical cavity using the method described above. For $\tau>10$~s, where cavity drift would limit the measurement accuracy at optical frequencies, the RF output of the frequency comb is compared to a hydrogen maser and a cryogenic sapphire oscillator (CSO) \cite{Fluhr2023} by the three-cornered hat method \cite{Gray1974}. The Allan deviations are $(31.4, 16.5, 3.4)\times 10^{-15}/\sqrt{\tau}$ for SF, TF, and MF spectroscopy, respectively, reaching a floor of approximately $4\times 10^{-15}$. The apparatus is not temperature-stabilized and the interferometer optics are not enclosed, so we attribute this instability primarily to variations in optical phase within the interferometer caused by air currents and mechanical drift. Fringe phase shifts produce effective frequency shifts that are independent of fringe amplitude, so all three spectroscopy methods approach a common noise floor at long averaging times.

Coupling to off-resonant laser lines contributes a small frequency shift that depends on the laser modulation parameters. We estimate the total fractional frequency shift from the measured laser spectrum (Fig.~\ref{fig:fig3}(b), bottom panel) to be of order $5\times 10^{-15}$, comparable to the second-order Doppler and Zeeman shifts for the current experimental parameters. However, the contribution to long-term instability should be minimal since the spectrum can be measured directly and the shift calculated.

In conclusion, we have demonstrated high-stability optical clock operation using a laser-cooled atomic beam, addressing an order-of-magnitude higher atom number by modulating the laser to perform spectroscopy simultaneously on many transverse velocity classes. Allan deviations measured at one second averaging time are competitive with trapped-atom optical clocks and the measured flux supports a QPN-limited stability of order $10^{-17}/\sqrt{\tau}$. This work establishes multifrequency spectroscopy as a viable method for efficient narrow-line interrogation of thermal sources, with possible future applications in two-photon vapor cell clocks \cite{Martin2018,Zhang2015} and atom interferometers \cite{Biedermann2017}.

Previous experiments suggest straightforward methods for improving the clock performance. The interferometer optics could be enclosed in vacuum and constructed as a monolithic element in order to eliminate optical path length variations \cite{Olson2019}. Certain fringe phase fluctuations could also be eliminated using ``k-reversal", where the laser propagation direction is reversed periodically and the frequency correction is derived from the average of both directions \cite{Ito1994, Olson2019}. Finally, atomic beams typically pass through mechanical apertures to reduce the spread of transverse velocities; removing these constraints and using multifrequency spectroscopy could enable much larger atomic fluxes to be interrogated. With these improvements to the long-term performance, we anticipate cold-beam Ramsey-Bordé interferometers to be a competitive architecture for optical timekeeping. 

We are grateful to T. G. Akin, J. Whalen, K. Boyce, M.~M.~Beydler, and J. Hanssen for helpful discussions. This work is supported by Department of Navy Research, Development, Test, and Evaluation funding for the Clock Development program at the United States Naval Observatory.

\bibliography{notes.bib}
\end{document}